\documentclass[prd,aps,floats,floatfix,nofootinbib]{revtex4}
\usepackage{amsmath}
\usepackage{amsfonts}
\usepackage{graphicx}
\usepackage{slashed}
\usepackage{dcolumn}
\usepackage{bm}
\usepackage[dvips]{color}
\newcommand{\beq}{\begin{equation}}
\newcommand{\eeq}{\end{equation}}
\begin{document}
%%%%%%%%%%%%%%%%%
\newcommand{\vect}[1]{\overrightarrow{#1}}
\newcommand{\smbox}[1]{\mbox{\scriptsize #1}}
\newcommand{\tanbox}[1]{\mbox{\tiny #1}}
\newcommand{\vev}[1]{\langle #1 \rangle}
\newcommand{\Tr}[1]{\mbox{Tr}\left[#1\right]}
\newcommand{\cosb}{c_{\beta}}
\newcommand{\sinb}{s_{\beta}}
\newcommand{\tanb}{t_{\beta}}
\newcommand{\picwidth}{3.4in}
%%%%%%%%%%%%%%%%

\preprint{MSU-HEP-100614}
\title{Global Symmetries  and Renormalizability of Lee-Wick Theories}

\author{R. Sekhar Chivukula}
\email[]{sekhar@msu.edu}
\author{Arsham Farzinnia}
\email[]{farzinni@msu.edu}
\author{Roshan Foadi}
\email[]{foadiros@msu.edu}
\author{Elizabeth H. Simmons}
\email[]{esimmons@msu.edu}
\affiliation{Department of Physics,
Michigan State University, East Lansing, MI 48824, USA}
\date{\today}

\begin{abstract}

In this paper we discuss the global symmetries and the renormalizibility of Lee-Wick scalar 
QED. In particular, in the ``auxiliary-field" formalism we identify softly
broken $SO(1,1)$ global symmetries of the theory. We introduce $SO(1,1)$ invariant
gauge-fixing conditions that allow us to show in the auxiliary-field formalism directly
that the number of superficially divergent 
amplitudes in a LW Abelian gauge theory is finite.  To illustrate the renormalizability of
the theory, we explicitly carry out the one-loop renormalization
program in LW scalar QED and demonstrate how the counterterms required are 
constrained by the joint conditions of gauge- and $SO(1,1)$-invariance. 
We also compute the one-loop beta-functions
in LW scalar QED and contrast them with those of ordinary scalar QED.

\end{abstract}

\maketitle

%%%%%%%%%%%%%%%%%%%%%%%%%%%%%%%%%%%%%%
%%%%%%%%%%%%%%%%%%%%%%%%%%%%%%%%%%%%%%
%%%%%%%%%%%%%%%%%%%%%%%%%%%%%%%%%%%%%%

\section{Introduction}

Recently an extension of the standard model (SM)~\cite{Grinstein:2007mp} based on the ideas of Lee and Wick (LW)~\cite{Lee:1969fy,Lee:1970iw} has been proposed as a solution to the hierarchy problem. The LW SM features higher-derivative kinetic terms for each SM field. This gives rise to propagators that fall off with momentum more rapidly than the ordinary SM field propagators, thereby reducing the degree of divergence of loop diagrams. On the other hand higher covariant derivatives also introduce new momentum dependent interactions, which raise the degree of divergence of quantum fluctuations. Power counting arguments~\cite{Grinstein:2007mp} show that these two competing effects conspire to make all loop diagrams at most logarithimically divergent. If the scale associated with the higher derivative terms is of the order of the electroweak scale then the latter becomes stable against radiative corrections: no quadratic divergences are present at any order in perturbation theory, and no unnatural fine-tuning of parameters is required to hold the Higgs vacuum expectation value fixed at $v=246$ GeV.

The higher-derivative kinetic terms in the LW SM result in  propagators with more than one pole. In $N=2$ LW theories~\cite{Carone:2009it} there is only one higher derivative kinetic term for each field, corresponding to two-pole propagators.\footnote{We will focus on $N=2$ theories throughout this paper, though our results can be potentially generalized to arbitrary-$N$ LW theories.} In the $N=2$ LW SM the lighter pole corresponds to a SM state, while the heavier pole corresponds to a new LW ghost state with the same quantum numbers and negative norm. The poles associated with the LW fields lie on the physical sheet in the complex $p^2$ plane. This is dangerous, since the presence of ghosts in the {\em in} and {\em out} states of the $S$ matrix would lead to a violation of unitarity. In order to avoid this scenario, the LW ghosts must appear only as virtual states~\cite{Lee:1969fy}, so that the $S$ matrix elements are only built out of physical asymptotic states. Furthermore, the integration contour in momentum integrals involving ghost propagators must be modified so as to preserve unitarity~\cite{Cutkosky:1969fq}. The price to be paid for these modifications is the presence of unobservable acausal effects in scattering processes~\cite{Lee:1969fy,Grinstein:2008bg}. Phenomenological consequences and constraints on TeV scale LW ghosts have been studied in Refs~\cite{Rizzo:2007ae,Rizzo:2007nf,Alvarez:2009af,Krauss:2007bz,Alvarez:2008za,Underwood:2008cr,Carone:2009nu,Dulaney:2007dx,Carone:2008bs,Chivukula:2010nw}. 

The LW ghost fields can alternatively be described by switching to a ``auxiliary-field'' form of the theory~\cite{Grinstein:2007mp}, in which in addition to the SM fields there are ``LW fields'' with kinetic energy terms with the opposite sign from their SM counterparts. The opposite sign for the kinetic energy terms enforces the cancellations that soften the divergences in the theory. The main advantage of the auxiliary-field approach lies in the computation of loop diagrams, since aside from the overall sign the propagators are just ordinary propagators. 

In this paper we clarify two issues in the auxiliary-field description of the theory in the context of a simple, but non-trivial,
theory -- LW scalar quantum electrodynamics (QED).\footnote{Our analysis extends immediately to LW QED with
an arbitrary number of matter fields, either scalars or fermions.} First, the interaction terms involving the LW fields have a very particular form, which is not the most general one allowed by gauge invariance. For example, the couplings of the LW vector fields are identical to the gauge couplings of the corresponding SM gauge fields. This equality and others are essential if the cancellations softening or removing the infinities are to hold. On the one hand,  it is not clear why, a priori, this special form of the interactions should be preserved to all orders in perturbation theory. On the other hand, we know that it must be preserved, since power-counting shows that the equivalent higher derivative theory is free of quadratic divergences\footnote{In the auxiliary-field formulation power counting is more difficult, because of the cancellations involved between different diagrams.}. This note identifies approximate $SO(1,1)$ global symmetries of the auxiliary-field description of the theory that allow us to understand its structure. 

Second, we clarify the renormalizability of LW scalar QED in the presence of the 
massive ghost LW vector field. Because of the $q^\mu q^\nu/M_A^2$ term in a heavy vector
boson propagator (where $M_A$ is the heavy vector mass), power-counting in the auxiliary-field formalism is difficult.
However, the LW vector field is not quite a non-gauge vector field, since it corresponds to a massive pole in the propagator of a higher-derivative gauge field. We will identify two $SO(1,1)$ symmetric gauge fixing conditions that simplify the auxiliary-field LW analysis. In one case (``ordinary"), the gauge fixing forces the  $q^\mu q^\nu/M_A^2$ terms to appear with canceling signs in the gauge-LW propagator matrix.  In the other case (``no-mixing"),  the gauge-fixing eliminates the $q^\mu q^\nu/M_A^2$ term in the vector field propagators. Working in the no-mixing gauge allows us to show that the number of superficially divergent amplitudes in an Abelian gauge theory is finite, and the theory is therefore renormalizable. 

Finally, to illustrate the renormalizability of LW scalar QED, we explicitly carry out the one-loop renormalization
program and demonstrate how the counterterms required are constrained by the joint conditions of gauge-
and $SO(1,1)$-invariance. As a byproduct of these discussions, we compute the one-loop beta-functions
in LW scalar QED and contrast them with those of ordinary scalar QED.

In Sec.~\ref{sec:phi4} we introduce and illustrate the $SO(1,1)$ symmetries of a LW theory in the context of
$\phi^4$ theory. In Sec.~\ref{sec:lgrng} we consider LW scalar QED and derive the equivalent auxiliary-field description. We then analyze the global symmetries of the theory and explain how these protect the form of the Lagrangian against radiative corrections. In Sec.~\ref{sec:fixing} we show how gauge fixing can be implemented in an $SO(1,1)$ invariant fashion and derive the corresponding propagators. In Sec.~\ref{sec:renormalization} we show that the number of superficially divergent amplitudes is finite, and the theory is therefore renormalizable. Then we illustrate these results at one loop by carrying
out the renormalization program and computing the $\beta$-functions. Finally in Sec.~\ref{sec:conclusions} we offer our conclusions, and we sketch why a modified approach is needed for the case of non-Abelian gauge theories.

%%%%%%%%%%%%%%%%%%%%%%%%%%%%%%%%%%%%%%
%%%%%%%%%%%%%%%%%%%%%%%%%%%%%%%%%%%%%%
%%%%%%%%%%%%%%%%%%%%%%%%%%%%%%%%%%%%%%

\section{Lee-Wick $\phi^4$ Theory}
\label{sec:phi4}

We first consider LW $\phi^4$ theory for a complex scalar field in order
to introduce the auxiliary-field formalism and the $SO(1,1)$ global symmetry of the model, as well as to set our notational conventions.\footnote{In this paper we follow closely the conventions of 
Ref. \protect\cite{Grinstein:2007mp}.} Lee-Wick $\phi^4$ theory is defined by the higher-derivative Lagrangian
\begin{equation}
{\cal L}_{hd} = |\partial_\mu \hat{\phi}|^2-\frac{1}{\hat{M}^2}|\partial^2 \hat{\phi}|^2
-\hat{m}^2|\hat{\phi}|^2 -\frac{\hat{\lambda}}{4}|\hat{\phi}|^4~,
\label{eq:hdphi4}
\end{equation}
where $\hat{\phi}$ is a complex scalar field, and the Lee-Wick scale $\hat{M}$ parameterizes the energy at which the model deviates substantially from the standard $\phi^4$ model. As we will see, $\hat{M}$ also characterizes the mass
scale of the LW ghosts, so long as $\hat{m} \ll \hat{M}$. 
This Lagrangian is equivalent to one in which we introduce an ``auxiliary" complex
scalar field $\tilde{\phi}^\prime$ (the reason for the ``prime" will become clear in what follows)
\begin{equation}
{\cal L} = |\partial_\mu \hat{\phi}|^2 + \hat{M}^2 |\tilde{\phi}^\prime|^2 + 
\partial_\mu \hat{\phi} \partial^\mu \tilde{\phi}^{\prime \ast}
+\partial_\mu \hat{\phi}^* \partial^\mu \tilde{\phi}^\prime-\hat{m}^2|\hat{\phi}|^2 -\frac{\hat{\lambda}}{4}|\hat{\phi}|^4~.
\label{eq:auxphi4}
\end{equation}
Making the change of variable
\begin{equation}
\hat{\phi} = \phi^\prime - \tilde{\phi}^\prime~,
\end{equation}
we find
\begin{equation}
{\cal L} = 
|\partial_\mu \phi^\prime|^2-|\partial_\mu \tilde{\phi}^\prime|^2 +\hat{M}^2 |\tilde{\phi}^\prime|^2 
-\hat{m}^2|\phi^\prime-\tilde{\phi}^\prime|^2 - \frac{\hat{\lambda}}{4} |\phi^\prime-\tilde{\phi}^\prime|^4~.
\end{equation}
The symplectic rotation
\begin{equation}
\left(\begin{array}{c} \phi^\prime \\ \tilde{\phi}^\prime \end{array}\right)
=\left(\begin{array}{cc}\cosh\theta & \sinh\theta \\ \sinh\theta & \cosh\theta \end{array}\right)
\left(\begin{array}{c} \phi \\ \tilde{\phi} \end{array}\right) \ ,
\label{eq:symplectic}
\end{equation}
where
\begin{eqnarray}
\tanh 2\theta = \frac{-2\hat{m}^2/\hat{M}^2}{1-2\hat{m}^2/\hat{M}^2} \ ,
\end{eqnarray}
diagonalizes the scalar field mass terms while preserving the symplectic structure of the kinetic terms~\cite{Grinstein:2007mp}. Hence we arrive at the auxiliary-field description of the LW $\phi^4$ theory
\begin{eqnarray}
{\cal L}_{\phi^4} &=& \  |\partial_\mu \phi|^2 - |\partial_\mu \tilde{\phi}|^2 +M^2 |\tilde{\phi}|^2 -m^2|\phi|^2 
- \frac{\lambda}{4} |\phi-\tilde{\phi}|^4
=\  |\partial_\mu \phi|^2 - |\partial_\mu \tilde{\phi}|^2 +M^2 |\tilde{\phi}|^2 -m^2|\phi|^2 \nonumber \\
&-& \frac{\lambda}{4} |\phi|^4 + \frac{\lambda}{2} |\phi|^2\left(\phi\tilde{\phi}^\ast+\phi^\ast\tilde{\phi}\right)
-  \lambda |\phi|^2 |\tilde{\phi}|^2 - \frac{\lambda}{4} \left(\phi^2\tilde{\phi}^{\ast 2}+\phi^{\ast 2}\tilde{\phi}^2\right)
+ \frac{\lambda}{2} |\tilde{\phi}|^2\left(\phi\tilde{\phi}^\ast+\phi^\ast\tilde{\phi}\right)
- \frac{\lambda}{4} |\tilde{\phi}|^4~,
\label{eq:phi4}
\end{eqnarray}   
where
\begin{eqnarray}
M^2 & = & \cosh^2\theta\ \hat{M}^2-e^{-2\theta}\ \hat{m}^2 \nonumber \\
m^2 & = & e^{-2\theta}\ \hat{m}^2 - \sinh^2\theta\ \hat{M}^2 \nonumber \\
\lambda & = & e^{-4\theta}\ \hat{\lambda} \ .
\label{eq:relabel}
\end{eqnarray}
Note that the kinetic term of the $\tilde{\phi}$ field has the opposite sign to the usual one, and hence
the corresponding particle has negative norm and is the LW ghost field. Furthermore, the mass of
the LW ghost $M$ is, in the limit $\hat{m} \ll \hat{M}$, approximately the LW scale $\hat{M}$
introduced in Eq. (\ref{eq:hdphi4}).

This  theory has an exact global $U(1)$ symmetry, but is not the most general $U(1)$ symmetric renormalizable Lagrangian that can be built out of the ordinary field $\phi$ and the ghost field $\tilde{\phi}$ charged under the $U(1)$ symmetry. In particular, the six interaction terms in the second line can in principle have six independent couplings. 
However the dimension-four terms in Eq.~(\ref{eq:phi4}) do have an additional 
$SO(1,1)$ symmetry, under which the fields transform as
\begin{equation}
\left(\begin{array}{c} \phi \\ \tilde{\phi} \end{array}\right)
\to\left(\begin{array}{cc}\cosh\beta & \sinh\beta \\ \sinh\beta & \cosh\beta \end{array}\right)
\left(\begin{array}{c} \phi \\ \tilde{\phi} \end{array}\right)~, 
\end{equation}
so long as we also promote $\lambda$ to a spurion field that
transforms as
\begin{equation}
\quad \lambda\to e^{4\beta}\ \lambda \ .
\end{equation}

Furthermore, the Lagrangian of Eq.~(\ref{eq:phi4}) is the most general renormalizable and $U(1)$-symmetric Lagrangian with $SO(1,1)$-symmetric dimension-four terms. The different mass terms for $\phi$ and $\tilde{\phi}$ break the $SO(1,1)$
symmetry, but do so only softly. They are also the only $U(1)$-preserving soft terms that break $SO(1,1)$. Thus in the 
LW $\phi^4$ theory, the global $SO(1,1)$ symmetry of the dimension-four terms implies that loop corrections can only modify the structure of the mass terms, introducing a mixing term between $\phi$ and $\tilde{\phi}$ with infinite coefficient. This can always be diagonalized via a symplectic rotation (of the form given in Eq. (\ref{eq:symplectic})), which leaves the rest of the Lagrangian unchanged, except for a redefinition of the coupling. Hence  Lee-Wick $\phi^4$ theory is renormalizable by power-counting.

LW $\phi^4$ theory  is rather simple, because aside from mass renormalization the theory is finite. The LW scenario is however much less trivial in LW gauge theories, because of the new momentum dependent interactions in the higher derivative formulation. In this case global symmetries  are important to understand the full structure of the theory. In the following, we will show that Abelian $N=2$ LW theories have a softly broken $SO(1,1)^{m+1}$ symmetry, where $m$ is the number of matter fields, and the remaining $SO(1,1)$ transformation acts on the vector fields. Since the $SO(1,1)^{m+1}$ breaking is soft, the special relation between the LW couplings and the ordinary couplings is protected against radiative corrections.

%%%%%%%%%%%%%%%%%%%%%%%%%%%%%%%%%%%%%%
%%%%%%%%%%%%%%%%%%%%%%%%%%%%%%%%%%%%%%
%%%%%%%%%%%%%%%%%%%%%%%%%%%%%%%%%%%%%%

\section{Global Symmetries of Lee-Wick Scalar QED}\label{sec:lgrng}

Let us now study an N=2 LW theory of scalar electrodynamics.  In the higher-derivative formulation, the Lagrangian is\footnote{In non-Abelian theories there can be additional dimension-six higher-derivative operators, which lead to heavy vector scattering amplitudes growing like $E^2$, where $E$ is the center-of-mass energy~\cite{Grinstein:2007iz}. For $N\geq 2$ LW theories see, for example, Ref.~\cite{Carone:2008iw}.}
\begin{eqnarray}
{\cal L}_{\rm hd} = -\frac{1}{4}\hat{F}_{\mu\nu}^2
+\frac{1}{2 M_A^2} \left(\partial^\mu \hat{F}_{\mu\nu}\right)^2
+|\hat{D}_\mu \hat{\phi}|^2 - \frac{1}{\hat{M}^2} |\hat{D}^2 \hat{\phi}|^2
-\hat{m}^2|\hat{\phi}|^2 - \frac{\hat{\lambda}}{4} |\hat{\phi}|^4 \ ,
\label{eq:higher}
\end{eqnarray}
where
\begin{eqnarray}
\hat{D}_\mu \equiv \partial_\mu  - i\ g \ \hat{A}_\mu  \ .
\label{eq:covariant}
\end{eqnarray}
The scalar sector is simply that of $\phi^4$ theory as shown in Eq. (\ref{eq:hdphi4}), and hence
our analysis of this Lagrangian will parallel the discussion of Sec. {\ref{sec:phi4}.
Introducing auxiliary fields, now for both the vector and the scalar, and using
the notation described above, we see that the Lagrangian of Eq.~(\ref{eq:higher}) is equivalent to
\begin{eqnarray}
{\cal L} = -\frac{1}{4}\hat{F}_{\mu\nu}^2 - \partial^\mu \tilde{A}^\nu\  \hat{F}_{\mu\nu} -\frac{M_A^2}{2} \tilde{A}_\mu^2
+|\hat{D}_\mu \hat{\phi}|^2 +\hat{M}^2 |\tilde{\phi}^\prime|^2 + \hat{D}_\mu\hat{\phi} \hat{D}^\mu\tilde{\phi}^{\prime\ast}
+ \hat{D}_\mu\hat{\phi}^\ast \hat{D}^\mu\tilde{\phi}^\prime 
-\hat{m}^2|\hat{\phi}|^2 - \frac{\hat{\lambda}}{4} |\hat{\phi}|^4 \ ,
\label{eq:lower1}
\end{eqnarray}
to all orders in perturbation theory. 
%In fact ${\cal L}$ is quadratic in $\tilde{A}_\mu$ and $\tilde{\phi}^\prime$, so we can perform the functional integral exactly by just completing the square. This clearly gives back ${\cal L}_{\rm hd}$. 
Changing variables from $\hat{A}_\mu$, $\tilde{A}_\mu$, $\hat{\phi}$, $\tilde{\phi}^\prime$ to $A_\mu$, $\tilde{A}_\mu$, $\phi^\prime$, $\tilde{\phi}^\prime$, where
\begin{eqnarray}
\hat{A}_\mu & = & A_\mu - \tilde{A}_\mu \ , \label{eq:AAtilde} \\
\hat{\phi} & = & \phi^\prime - \tilde{\phi}^\prime \ ,
\end{eqnarray}
and substituting in Eqs.~(\ref{eq:covariant}) and (\ref{eq:lower1}), gives
\begin{eqnarray}
{\cal L} &=& -\frac{1}{4}F_{\mu\nu}^2 + \frac{1}{4}\tilde{F}_{\mu\nu}^2 -\frac{M_A^2}{2} \tilde{A}_\mu^2
+|D_\mu \phi^\prime|^2-|D_\mu \tilde{\phi}^\prime|^2 +\hat{M}^2 |\tilde{\phi}^\prime|^2 
-\hat{m}^2|\phi^\prime-\tilde{\phi}^\prime|^2 - \frac{\hat{\lambda}}{4} |\phi^\prime-\tilde{\phi}^\prime|^4 \ \nonumber \\
& - & i\ g\ \tilde{A}_\mu \left(\phi^\prime D^\mu \phi^{\prime\ast} - \phi^{\prime\ast} D^\mu \phi^\prime \right)
+ i\ g\ \tilde{A}_\mu \left(\tilde{\phi}^\prime D^\mu \tilde{\phi}^{\prime\ast} - \tilde{\phi}^{\prime\ast} D^\mu \tilde{\phi}^\prime \right)
+g^2 \tilde{A}_\mu^2\left(|\phi^\prime|^2-|\tilde{\phi}^\prime|^2\right) \ ,
\end{eqnarray}
where now the covariant derivative is in terms of $A_\mu$,
\begin{eqnarray}
D_\mu = \partial_\mu- i\ g \ A_\mu \ .
\end{eqnarray}
The symplectic rotation
of Eq. (\ref{eq:symplectic}) again diagonalizes the scalar field mass terms while preserving the symplectic structure of the kinetic terms~\cite{Grinstein:2007mp}. Since the gauge interactions stem from kinetic terms, and the $\phi^4$-interaction has a symplectic structure as well, it follows that Eq.~(\ref{eq:symplectic}) only diagonalizes the mass terms leaving the rest of the Lagrangian invariant in form. In terms of $\phi$ and $\tilde{\phi}$ the Lagrangian now reads
\begin{eqnarray}
{\cal L} &=& -\frac{1}{4}F_{\mu\nu}^2 + \frac{1}{4}\tilde{F}_{\mu\nu}^2 -\frac{M_A^2}{2} \tilde{A}_\mu^2
+|D_\mu \phi|^2-|D_\mu \tilde{\phi}|^2 +M^2 |\tilde{\phi}|^2 
-m^2|\phi|^2 - \frac{\lambda}{4} |\phi-\tilde{\phi}|^4 \ \nonumber \\
& + & i\ g\ \tilde{A}_\mu \left(\phi\ D^\mu \phi^\ast - \phi^\ast D^\mu \phi \right)
- i\ g\ \tilde{A}_\mu \left(\tilde{\phi}\ D^\mu \tilde{\phi}^\ast - \tilde{\phi}^\ast D^\mu \tilde{\phi} \right)
+g^2 \tilde{A}_\mu^2\left(|\phi|^2-|\tilde{\phi}|^2\right) \ ,
\label{eq:Ltwofields}
\end{eqnarray}
where we redefine parameters as in Eq. (\ref{eq:relabel}).

The Lagrangian of Eq.~(\ref{eq:Ltwofields}) has an exact $U(1)$ gauge symmetry. In the limit $\lambda\to 0$ the global symmetry is promoted to $U(1)\times U(1)$, because the $\phi$ and $\tilde{\phi}$ fields can now rotate independently, and only the diagonal $U(1)$ subgroup is gauged. Thus we expect loop corrections to generate $U(1)$-symmetric terms -- some with infinite coefficients --  that will be $U(1)\times U(1)$-symmetric in the $\lambda\to 0$ limit. Eq.~(\ref{eq:Ltwofields}) is not the most general renormalizable Lagrangian with this symmetry structure; for example, the coefficients of the interactions involving $\tilde{A}_\mu$ could be arbitrary. Notice, however, that this Lagrangian can be re-arranged in the form
\begin{eqnarray}
{\cal L} &=& -\frac{1}{4}F_{\mu\nu}^2 + \frac{1}{4}\tilde{F}_{\mu\nu}^2 -\frac{M_A^2}{2} \tilde{A}_\mu^2
+|\partial_\mu \phi|^2-|\partial_\mu \tilde{\phi}|^2 +M^2 |\tilde{\phi}|^2
-m^2|\phi|^2 - \frac{\lambda}{4} |\phi-\tilde{\phi}|^4 \ \nonumber \\
& - & i\ g\ (A_\mu-\tilde{A}_\mu) \left(\phi\ \partial^\mu \phi^\ast- \tilde{\phi}\ \partial^\mu \tilde{\phi}^\ast - {\rm h.c.}\right)
-g^2 (A_\mu-\tilde{A}_\mu)^2\left(|\phi|^2-|\tilde{\phi}|^2\right) \ .
\label{eq:Ltwofields2}
\end{eqnarray}
In the limit $M_A\to 0$, and treating the gauge coupling as a spurion field, the Lagrangian respects a global $SO(1,1)$ symmetry under which
\begin{equation}
\left(\begin{array}{c} A_\mu \\ \tilde{A}_\mu \end{array}\right)
\to\left(\begin{array}{cc}\cosh\alpha & \sinh\alpha \\ \sinh\alpha & \cosh\alpha \end{array}\right)
\left(\begin{array}{c} A_\mu \\ \tilde{A}_\mu \end{array}\right), \quad g \to e^{\alpha}\ g \ .
\label{eq:SOgauge}
\end{equation}
As mentioned in Section II, an additional $SO(1,1)$ global symmetry for the scalar field arises in the limit $M\to m$, when $\lambda$ is treated as a spurion field:
\begin{equation}
\left(\begin{array}{c} \phi \\ \tilde{\phi} \end{array}\right)
\to\left(\begin{array}{cc}\cosh\beta & \sinh\beta \\ \sinh\beta & \cosh\beta \end{array}\right)
\left(\begin{array}{c} \phi \\ \tilde{\phi} \end{array}\right), \quad \lambda\to e^{4\beta}\ \lambda \ .
\label{eq:SOscalar}
\end{equation}

In Sec.~\ref{sec:renormalization} we will argue that this theory is renormalizable, because the number of superficially divergent amplitudes is finite, and no operators of dimension greater than four are present in the auxiliary-field formulation. We may wonder whether radiative corrections require introducing dimension-four $SO(1,1)\times SO(1,1)$-breaking counterterms. However, as in the LW $\phi^4$ theory described above, the answer is no: since $SO(1,1)\times SO(1,1)$ is only softly broken by mass terms,\footnote{The renormalizability of massive Abelian gauge theory
\protect\cite{Kroll:1967it} -- which arises from the fact that the Abelian gauge boson couples
to a conserved current -- insures that the gauge-boson mass term is ``soft".} the $SO(1,1)\times SO(1,1)$-breaking corrections to the renormalizable terms are finite. 
Furthermore Eq.(\ref{eq:Ltwofields2}) is the most general $U(1)$ gauge Lagrangian with dimension-four $SO(1,1)\times SO(1,1)$-symmetric terms. Since renormalizability prevents higher dimensional operators from being generated, we conclude that the form of the Lagrangian is protected to all orders against radiative corrections, with the exception of the scalar field mass terms. However, as we have already seen, these can be diagonalized with a symplectic rotation, without affecting the rest of the Lagrangian. In the simple example we have shown there is only one matter field: for an arbitrary number $m$ of matter fields the global symmetry is promoted to $SO(1,1)\times SO(1,1)^m$, since each field is acted on with a different $SO(1,1)$ symmetry transformation, and the conclusions about renormalizability persist, {\it mutatis mutandis}.

%%%%%%%%%%%%%%%%%%%%%%%%%%%%%%%%%%%%%%
%%%%%%%%%%%%%%%%%%%%%%%%%%%%%%%%%%%%%%
%%%%%%%%%%%%%%%%%%%%%%%%%%%%%%%%%%%%%%

\section{Gauge Fixing}\label{sec:fixing}

In order to quantize the electromagnetic field, one must introduce a gauge-fixing term. To facilitate our subsequent analyses of divergences and renormalizability, we will find it most convenient to employ gauge-fixing functions that respect the $SO(1,1)$ symmetry; otherwise it can be unnecessarily difficult to recognize when significant cancellations occur\footnote{For example, Ref.~\cite{Grinstein:2007mp} employs an $SO(1,1)$ violating gauge-fixing term $-\frac{1}{2\xi}\left(\partial^\mu A_\mu\right)^2$
which leads to diagonal propagators of the form
\begin{eqnarray}
P_{AA}^{\mu\nu} = \frac{-i}{q^2}\left[g^{\mu\nu}-(1-\xi)\frac{q^\mu q^\nu}{q^2}\right], \ \ 
P_{\tilde{A}\tilde{A}}^{\mu\nu} = \frac{i}{q^2-M_A^2}\left[g^{\mu\nu}-\frac{q^\mu q^\nu}{M_A^2}\right] \ . \nonumber
\end{eqnarray}
Because the $q^\mu q^\nu/M_A^2$ term is only present in the LW-photon propagator, there is no simple cancellation of the badly-behaved terms and the theory appears to suffer from quadratic divergences and nonrenormalizability at one loop.   The reason that Ref.~\cite{Grinstein:2007mp} found no quadratic divergences when computing the self energy amplitudes for a massless scalar field at zero momentum is that the quadratic divergence vanishes in the limit $m\to 0$ and $q\to 0$, since it is necessarily of the form $\Lambda^2 m^2/M_A^2$ or $\Lambda^2 q^2/M_A^2$. These quadratic divergences are ``gauge-artifacts" \protect\cite{Grinstein:2008qq} in the sense that they contribute to both scalar wavefunction and mass renormalization in such a way that the pole-mass of the scalar is {\it not} quadratically sensitive to the cutoff.}.  To ensure that the symmetry will be preserved, it is sufficient to write the gauge fixing term in terms of $\hat{A}_\mu/\sqrt{\xi}$, where $\xi$ is the gauge fixing parameter. If one treats $\xi$ as a spurion field, the $SO(1,1)$ transformation
\begin{eqnarray}
\hat{A}_\mu \to e^{-\alpha}\ \hat{A}_\mu, \quad \xi \to e^{-2\alpha}\ \xi
\end{eqnarray}
clearly leaves $\hat{A}_\mu/\sqrt{\xi}$ invariant. We will consider two different $SO(1,1)$ symmetric gauge-fixing scenarios that are each convenient in different circumstances, and will denote them as ``ordinary" and ``no-mixing" gauge fixing.

\subsection{Ordinary Gauge Fixing}

First, let us consider a gauge fixing function of the typical form $G(\hat{A})=\partial^\mu \hat{A}_\mu$. In $R_\xi$ gauge this amounts to adding to the Lagrangian the $SO(1,1)$ symmetric gauge-fixing term
\begin{eqnarray}
{\cal L}^{\rm ordinary}_{\rm fixing} = -\frac{1}{2\xi}\left(\partial^\mu \hat{A}_\mu\right)^2 \ .
\end{eqnarray}
Using Eq.~(\ref{eq:AAtilde}) to rewrite this in terms of $\tilde{A}^\mu$ and $A^\mu$, one obtains 
\begin{eqnarray}
{\cal L}^{\rm ordinary}_{\rm fixing} = -\frac{1}{2\xi}\left(\partial^\mu A_\mu\right)^2 - \frac{1}{2\xi}\left(\partial^\mu \tilde{A}_\mu\right)^2
+\frac{1}{\xi}\partial^\mu A_\mu\ \partial^\nu \tilde{A}_\nu.
\end{eqnarray}
With the gauge fixing included, and after integrating by parts, the gauge field Lagrangian reads
\begin{eqnarray}
{\cal L}^{\rm ordinary}_{\rm gauge} = \frac{1}{2}A_\mu \left[g^{\mu\nu}\partial^2-(1-1/\xi)\partial^\mu\partial^\nu\right]A_\nu
-\frac{1}{2}\tilde{A}_\mu \left[g^{\mu\nu}(\partial^2-M_A^2)-(1+1/\xi)\partial^\mu\partial^\nu\right]\tilde{A}_\nu
- A_\mu \frac{1}{\xi} \partial^\mu \partial^\nu \tilde{A}_\nu \ .
\end{eqnarray}

We can invert the diagonal terms, in momentum space, to find the partial propagators
\begin{eqnarray}
D_{AA}^{\mu\nu} = \frac{-i}{q^2}\left[g^{\mu\nu}-(1-\xi)\frac{q^\mu q^\nu}{q^2}\right], \ \ 
D_{\tilde{A}\tilde{A}}^{\mu\nu} = \frac{i}{q^2-M_A^2}\left[g^{\mu\nu}-(1+\xi)\frac{q^\mu q^\nu}{q^2+\xi M_A^2}\right] \ .
\end{eqnarray}
Then the full tree-level photon and LW-photon propagators, as well as the mixed propagators, can be computed by resumming the Dyson series to obtain
\begin{eqnarray}
P_{AA}^{\mu\nu} = \frac{-i}{q^2}\left[g^{\mu\nu}-(1-\xi)\frac{q^\mu q^\nu}{q^2} + \frac{q^\mu q^\nu}{M_A^2}\right], \ \ 
P_{\tilde{A}\tilde{A}}^{\mu\nu} = \frac{i}{q^2-M_A^2}\left[g^{\mu\nu}-\frac{q^\mu q^\nu}{M_A^2}\right], \ \ 
P_{\tilde{A}A}^{\mu\nu} = P_{A\tilde{A}}^{\mu\nu} = -\frac{i\ q^\mu q^\nu}{q^2 M_A^2} \ .
\label{eq:proplower}
\end{eqnarray}
Notice that only the photon propagator depends on the gauge-fixing parameter $\xi$, since the photon is the only true gauge field in this theory. 

Up to an overall sign, the $\tilde{A}_\mu$ propagator is identical to the unitary-gauge propagator of a massive gauge boson, in a spontaneously broken gauge theory. In particular, it contains the $q^\mu q^\nu/M_A^2$ term which would apparently render  the theory nonrenormalizable and reintroduce quadratic divergences. However the photon propagator and the mixed propagators contain the same term; when the $P_{AA}$, $P_{\tilde{A}\tilde{A}}$, and $P_{A\tilde{A}}$ propagators are all included in loop integrals, the badly-behaved terms cancel, and quadratic divergences are avoided. This can be seen even more clearly from the form of the $P^{\mu\nu}_{\hat{A}\hat{A}}$ propagator.  Recalling that all gauge interactions depend on $A_\mu-\tilde{A}_\mu=\hat{A}_\mu $ and working in Feynman gauge  ($\xi = 1$), we obtain
\begin{eqnarray}
P^{\mu\nu}_{\hat{A}\hat{A}} &=& P^{\mu\nu}_{AA}+P^{\mu\nu}_{\tilde{A}\tilde{A}} - 2 P^{\mu\nu}_{A\tilde{A}}
=\frac{-i}{q^2-q^4/M_A^2}\left[g^{\mu\nu}-\frac{q^\mu q^\nu}{M_A^2}\right] \ ,
\label{eq:hatprop}
\end{eqnarray}
which decays like $1/q^2$ for large values of the momentum.

\subsection{No-Mixing Gauge Fixing}

Next, we consider the alternative\footnote{In non-Abelian theories this introduces a $q^2/M_A^2$ expansion in the gauge-ghost interaction, which renders it less interesting. In Abelian gauge theories, however, the ghosts are decoupled. } gauge-fixing function $G(\hat{A})=\left(1+\partial^2/M_A^2\right)^{1/2}\partial^\mu \hat{A}_\mu$. The resulting $SO(1,1)$-symmetric gauge-fixing Lagrangian is
\begin{eqnarray}
{\cal L}^{\rm no-mixing}_{\rm fixing} = -\frac{1}{2\xi}\left(\partial^\mu \hat{A}_\mu\right)^2
+\frac{1}{2 \xi M_A^2} \left(\partial^\mu\partial^\nu \hat{A}_\nu \right)^2 \ .
\end{eqnarray}
Adding this to the original higher-derivative gauge Lagrangian gives, after integration by parts,
\begin{eqnarray}
{\cal L}^{\rm no-mixing}_{\rm gauge} = \frac{1}{2}\hat{A}_\mu \left[g^{\mu\nu}\partial^2-(1-1/\xi)\partial^\mu\partial^\nu\right]
\left(1+\partial^2/M_A^2\right)\hat{A}_\nu \ .
\label{eq:no-mixing-hd-gauge}
\end{eqnarray}
This is equivalent to
\begin{eqnarray}
{\cal L}^{\rm no-mixing}_{\rm gauge} = \frac{1}{2}\hat{A}_\mu \left[g^{\mu\nu}\partial^2-(1-1/\xi)\partial^\mu\partial^\nu\right]\hat{A}_\nu
+\hat{A}_\mu \left[g^{\mu\nu}\partial^2-(1-1/\sqrt{\xi})\partial^\mu\partial^\nu\right]\tilde{A}_\nu -\frac{M_A^2}{2}\tilde{A}_\mu^2 \ ,
\label{eq:no-mixing-gauge}
\end{eqnarray}
in the sense that solving the equations of motion for $\tilde{A}_\mu$ and inserting the solution in (\ref{eq:no-mixing-gauge}) recovers the form of (\ref{eq:no-mixing-hd-gauge}).  

At this point, we can eliminate the $\hat{A}^\mu$ field from Eq.~(\ref{eq:no-mixing-gauge}) via  Eq.~(\ref{eq:AAtilde}); the Lagrangian will include both diagonal and mixing terms in $A^\mu$ and $\tilde{A}^\mu$, and the full tree-level propagators can, again, be computed by summing the Dyson series. However for $\xi=1$ the mixing term vanishes\footnote{As an alternative to the $\xi=1$ gauge, one could replace Eq.~(\ref{eq:AAtilde}) with
\begin{eqnarray}
\hat{A}_\mu = A_\mu - \left[\delta_\mu^\nu - (1-\sqrt{\xi})\frac{q_\mu q^\nu}{q^2}\right]\tilde{A}_\nu \ , \nonumber
\end{eqnarray}
in momentum space. This cancels the off-diagonal terms for any value of $\xi$, at the price of introducing non-local interactions in coordinate space for $\xi\neq1$. }, and we obtain the simpler, diagonal Lagrangian 
\begin{eqnarray}
{\cal L}_{\rm gauge, \xi=1} = \frac{1}{2}A_\mu \partial^2 A^\mu
-\frac{1}{2}\tilde{A}_\mu (\partial^2-M_A^2) \tilde{A}^\mu \ .
\end{eqnarray}
The corresponding propagators
\begin{eqnarray}
P_{AA}^{\mu\nu} = \frac{-i\ g^{\mu\nu}}{q^2}, \ \ 
P_{\tilde{A}\tilde{A}}^{\mu\nu} = \frac{i\ g^{\mu\nu}}{q^2-M_A^2} \ ,
\label{eq:prophigher}
\end{eqnarray}
have no $q^\mu q^\nu$ terms so they are well-behaved at high energies.  This is equally clear if we construct the $\hat{A}$ propagator,
\begin{eqnarray}
P^{\mu\nu}_{\hat{A}\hat{A}} &=& P^{\mu\nu}_{AA}+P^{\mu\nu}_{\tilde{A}\tilde{A}} - 2 P^{\mu\nu}_{A\tilde{A}}
=\frac{-i g^{\mu\nu}}{q^2-q^4/M_A^2} \ ,
\label{eq:hatprophigher}
\end{eqnarray}
which falls off like $q^{-4}$ in the ultraviolet.

%Notice, however, that even for $\xi\neq 1$ the $\tilde{A}_\mu$ propagator has always the form of Eq.~(\ref{eq:prophigher}), independent %of $\xi$: in fact $\tilde{A}_\mu$ is not a gauge field, as the hard mass term is not gauge invariant.

\medskip
We will now use the two convenient gauges introduced in this section to explore LW scalar QED at one-loop.

%%%%%%%%%%%%%%%%%%%%%%%%%%%%%%%%%%%%%%
%%%%%%%%%%%%%%%%%%%%%%%%%%%%%%%%%%%%%%
%%%%%%%%%%%%%%%%%%%%%%%%%%%%%%%%%%%%%%

\section{One-Loop Renormalization}\label{sec:renormalization}

 We will start by establishing an upper bound on the superficial degree of divergence of Feynman diagrams in N=2 LW scalar QED.  For specificity, we work in the auxiliary-field formulation and employ the no-mixing  $\xi=1$ gauge. Recalling that each loop integral introduces four powers of momentum in the numerator, each trilinear gauge-scalar-scalar vertex introduces one power of momentum in the numerator, and the propagator has two powers of momentum in the denominator, we arrive at:
\begin{eqnarray}
D \leq 4L - 2P_A - 2P_{\tilde{A}} - 2P_\phi - 2P_{\tilde{\phi}} + V_{gss} \ .
\label{eq:divergence}
\end{eqnarray}
where $L$ is the number of loops, $P_f$ is the number of propagators of the $f$ field, and $V_{gss}$ is the number of trilinear gauge-scalar-scalar vertices. The number of loop integrals is, in turn, given by the total number of propagators (each carrying its own momentum space integral) minus the total number of vertices (each carrying a momentum-space delta function) plus one, since an overall delta function ensures momentum conservation for the external fields. Therefore, denoting the number of quartic gauge-gauge-scalar-scalar vertices by $V_{ggss}$ and the number of 4-point scalar vertices by $V_{ssss}$, we have
\begin{eqnarray}
L &=& P_A + P_{\tilde{A}} + P_\phi + P_{\tilde{\phi}} - V_{gss} - V_{ggss} - V_{ssss} + 1 \ .
\label{eq:elloop}
\end{eqnarray}
Finally we can relate the number of lines attached to a vertex to the number of propagators (each connecting two vertices) and the number of external lines $N_f$:
\begin{eqnarray}
V_{gss} &+& 2 V_{ggss} =  2P_A + 2P_{\tilde{A}} + N_A + N_{\tilde{A}}\ , \nonumber \\
2V_{gss} &+& 2V_{ggss} + 4 V_{ssss} = 2P_\phi + 2P_{\tilde{\phi}} + N_\phi + N_{\tilde{\phi}} \ ,
\label{eq:verte}
\end{eqnarray} 
where the first relation deals with gauge lines and the second with scalar lines. Inserting Eqs.~(\ref{eq:elloop}) and (\ref{eq:verte}) in Eq.~(\ref{eq:divergence}) yields
\begin{eqnarray}
D \leq 4 - N_A - N_{\tilde{A}} - N_{\phi} - N_{\tilde{\phi}} \ .
\label{eq:D}
\end{eqnarray}
This equation tells us that the number of superficially divergent amplitudes is finite; since no operators of dimension greater than four are present in the auxiliary-field Lagrangian, we conclude that the theory is renormalizable.

In order to confirm renormalizability explicitly and to verify how the $SO(1,1)\times SO(1,1)$ structure of the theory is protected against radiative corrections, we will now compute the infinite\footnote{If we were to compute the finite part as well, we would need to employ the Cutkosky-Landshoff-Olive-Polkinghorne prescription in order to avoid unitarity violation~\cite{Cutkosky:1969fq}. However the infinite part is not affected by this subtlety~\cite{Grinstein:2008qq}.} part of the divergent 1PI diagrams at one loop.   As a way of checking our results and exploring the detailed symmetry structure, we will compute the diagrams in both the ordinary and no-mixing gauges, with $\xi = 1$.  As we shall see, in the no-mixing  gauge only the vacuum polarization and self energy amplitudes are infinite, while in the ordinary gauge infinities also arise in the vertex corrections. Therefore, the way the counterterms preserve the symplectic structure of the theory is different in the two gauges.

\subsection{Counterterms}

Radiative corrections renormalize the fields and and mix the ordinary fields with the LW partners. This not only preserves the $U(1)$ gauge symmetry, but also does not generate any hard breaking of the global $SO(1,1)\times SO(1,1)$ symmetry. Let us derive the most general relation between bare and renormalized fields satisfying these requirements.  We will employ the standard QED nomenclature for the counterterms by using the subscript ``3'' for the photon wavefunction renormalization, ``2'' for the matter field wavefunction renormalization, and ``1'' for gauge vertex renormalization.

For the vector fields we have in general
\begin{eqnarray}
\left(\begin{array}{c} A^\mu \\ \tilde{A}^\mu \end{array}\right)
=\left(\begin{array}{cc} \sqrt{Z_3} & \sqrt{Z_3^\prime} \\ \sqrt{Z_3^{\prime\prime}} & \sqrt{\widetilde{Z}_3} \end{array}\right)
\left(\begin{array}{c} A_r^\mu \\ \tilde{A}_r^\mu \end{array}\right) \nonumber \ .
\end{eqnarray}
Gauge invariance requires $Z_3^{\prime\prime} = 0$, lest a photon mass term be generated. Preserving the form of the symplectic combination $A_\mu-\tilde{A}_\mu$ demands
\begin{eqnarray}
\sqrt{Z_3^\prime}=\sqrt{\widetilde{Z}_3}-\sqrt{Z_3} \nonumber \ .
\end{eqnarray}
Finally, substituting in the kinetic term Lagrangian and imposing the  $SO(1,1)$ symmetry on the counterterms gives
\begin{eqnarray}
\widetilde{Z}_3 = \frac{1}{Z_3} \ . \nonumber
\end{eqnarray}
Therefore the relation between bare and renormalized vector fields consistent with the symmetries of the theory is
\begin{eqnarray}
\left(\begin{array}{c} A^\mu \\ \tilde{A}^\mu \end{array}\right)
=\sqrt{Z_3}\left(\begin{array}{cc} 1 & \ \ \ Z_3^{-1}-1 \\ 0 & Z_3^{-1} \end{array}\right)
\left(\begin{array}{c} A_r^\mu \\ \tilde{A}_r^\mu \end{array}\right) \ , \quad Z_3\equiv 1+\delta_3 \ .
\label{eq:renormA}
\end{eqnarray}
Similarly, in order to preserve the $SO(1,1)$ symmetry on the scalar fields, the relation between bare and renormalized scalar fields must be a symplectic rotation times a wavefunction renormalization:
\begin{eqnarray}
\left(\begin{array}{c} \phi \\ \tilde{\phi} \end{array}\right)
= \sqrt{Z_2}\left(\begin{array}{cc}\cosh\eta & \sinh\eta \\ \sinh\eta & \cosh\eta \end{array}\right)
\left(\begin{array}{c} \phi_r \\ \tilde{\phi}_r \end{array}\right) \ , \quad Z_2\equiv 1+\delta_2 \ .
\label{eq:renormphi}
\end{eqnarray}

Substituting Eqs.~(\ref{eq:renormA}) and (\ref{eq:renormphi}) in the Lagrangian, Eq.~(\ref{eq:Ltwofields2}), and denoting the gauge-scalar-scalar vertex and gauge-gauge-scalar-scalar vertex renormalizations, respectively, by $Z_1 \equiv 1 + \delta_1$ and $Z_1^\prime = 1 + \delta_1^\prime$ leads to
\begin{eqnarray}
{\cal L} &=& -\frac{1}{4}F_{r\mu\nu}^2 + \frac{1}{4}\tilde{F}_{r\mu\nu}^2 -\frac{M_{Ar}^2}{2} \tilde{A}_{r\mu}^2
+|\partial_\mu \phi_r|^2-|\partial_\mu \tilde{\phi}_r|^2 +M_r^2 |\tilde{\phi}_r|^2
-m_r^2|\phi_r|^2 - \frac{\lambda_r}{4} |\phi_r-\tilde{\phi}_r|^4 \ \nonumber \\
& - & i\ g_r\ (A_{r\mu}-\tilde{A}_{r\mu}) \left(\phi_r\ \partial^\mu \phi_r^\ast- \tilde{\phi}_r\ \partial^\mu \tilde{\phi}_r^\ast - {\rm h.c.}\right)
-g_r^2 (A_{r\mu}-\tilde{A}_{r\mu})^2\left(|\phi_r|^2-|\tilde{\phi}_r|^2\right) + {\cal L}_{\rm ct} \ ,
\label{eq:Lrenorm}
\end{eqnarray}
where
\begin{eqnarray}
{\cal L}_{\rm ct} &=& -\frac{\delta_3}{4}F_{r\mu\nu}^2 + \frac{\delta_3}{2}F_{r\mu\nu} \tilde{F}_r^{\mu\nu}
-\frac{\delta_3}{4}\tilde{F}_{r\mu\nu}^2 -\frac{\delta_{M_A^2}}{2} \tilde{A}_{r\mu}^2 \nonumber \\
&+&\delta_2\ |\partial_\mu \phi_r|^2-\delta_2\ |\partial_\mu \tilde{\phi}_r|^2 +\delta_{M^2} |\tilde{\phi}_r|^2
-\delta_{m^2}|\phi_r|^2 -\delta_{mM}\left(\phi_r^\ast\tilde{\phi}_r+\phi_r\tilde{\phi}_r^\ast\right)
- \frac{\delta_\lambda}{4} |\phi_r-\tilde{\phi}_r|^4 \nonumber \\
&-& i\ \delta_1\ g_r\ (A_{r\mu}-\tilde{A}_{r\mu}) 
\left(\phi_r\ \partial^\mu \phi_r^\ast- \tilde{\phi}_r\ \partial^\mu \tilde{\phi}_r^\ast - {\rm h.c.}\right)
-\delta_1^\prime\ g_r^2 (A_{r\mu}-\tilde{A}_{r\mu})^2\left(|\phi_r|^2-|\tilde{\phi}_r|^2\right) \ .
\label{eq:Lct}
\end{eqnarray}

The renormalized trilinear and quartic gauge-scalar couplings are related to the bare couplings by
\begin{eqnarray}
g\ \sqrt{Z_3}\ Z_2 = g_r\ Z_1 \qquad\qquad g^2\ Z_3\ Z_2 = g_r^2\ Z_1^\prime\ ,
\end{eqnarray}
where gauge invariance guarantees
\begin{eqnarray}
Z_1 = Z_1^\prime = Z_2 
\label{eq:ward}
\end{eqnarray}
to all orders in perturbation theory. The renormalized mass parameters are related to the bare masses by
\begin{eqnarray}  
&& Z_2 \left[(\cosh\eta)^2\ m^2 - (\sinh\eta)^2\ M^2\right] = m_r^2 + \delta_{m^2} \nonumber \\
&& Z_2 \left[(\cosh\eta)^2\ M^2 - (\sinh\eta)^2\ m^2\right] = M_r^2 + \delta_{M^2} \nonumber \\
&& Z_2 (M^2-m^2)\cosh\eta\ \sinh\eta = -\ \delta_{mM} \ ,
\label{eq:ctrel}
\end{eqnarray}
whereas the renormalized and bare four-scalar couplings are related by
\begin{eqnarray}
\lambda\ Z_2^2\ e^{-4\eta} = \lambda_r + \delta_\lambda \ .
\end{eqnarray}
The vector field kinetic terms, in the counterterm Lagrangian, are now mixed. However it can be easily shown that these are still invariant under an $SO(1,1)$ transformation, provided that $\delta_3$ is promoted to a spurion field.

We shall now prove that this set of counterterms is sufficient to absorb all infinities at one loop. In the process, the $SO(1,1)\times SO(1,1)$ global symmetry leads to cancellation of the quadratic divergences in the scalar field self energy amplitudes. In order to simplify our notation we will drop the subscript $r$ everywhere, but it should be kept in mind that all fields and parameters involved in the calculations below are the renormalized ones.

\subsection{Vacuum Polarization Amplitudes}

\begin{figure}[!t]
\begin{center}
\includegraphics[width=5.2in,height=2.4in]{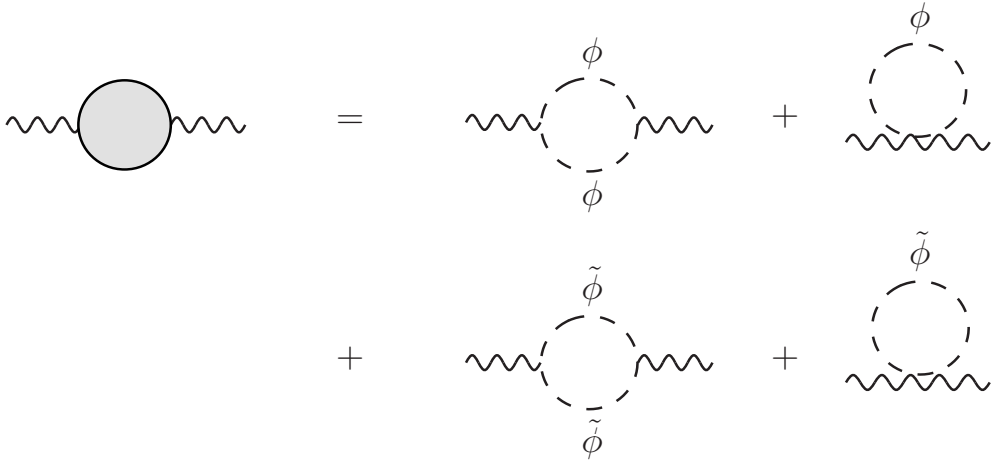}
\end{center}
\caption{One-loop contribution to the vacuum polarization amplitudes. Each external vector field is either a photon or a LW photon.}
\label{fig:VPA}
\end{figure}

We begin our examination of the infinite part of the divergent 1PI diagrams of LW scalar QED by computing the one-loop contributions to the vacuum polarization amplitudes for the vector fields.  The relevant diagrams are illustrated in Fig.~\ref{fig:VPA}, where each external field is either a photon or a LW photon. Since no gauge field propagators are involved in the one-loop diagrams, the results are manifestly gauge independent. We find
\begin{eqnarray}
i\ \Pi_{AA}^{\mu\nu}=i\ \Pi_{\tilde{A}\tilde{A}}^{\mu\nu}=-i\ \Pi_{A\tilde{A}}^{\mu\nu} =i\ \Pi(q^2)\ (q^2 g^{\mu\nu}-q^\mu q^\nu) \ ,
\label{eq:VPA}
\end{eqnarray}
where, in dimensional regularization,
\begin{eqnarray}
\Pi(q^2)= - 2\times \frac{e^2}{48\pi^2}\ \frac{1}{\epsilon}\ +\ {\rm finite\ terms} \ ,
\label{eq:Pi}
\end{eqnarray}
with $\epsilon\equiv 2-d/2$ as usual.  The explicit factor of two arises from the presence of the LW scalar loops, and the remaining factor is the ordinary scalar QED contribution. Since $\Pi_{\tilde{A}\tilde{A}}^{\mu\nu}$ contains no mass term, we have
\begin{eqnarray}
\delta_{M_A^2} = 0 \ .
\end{eqnarray}
The relevant counterterm contributions from the field-strength terms in Eq.~(\ref{eq:Lct}) are
\begin{eqnarray}
i\ \delta\Pi_{AA}^{\mu\nu}=i\ \delta\Pi_{\tilde{A}\tilde{A}}^{\mu\nu}
=-i\ \delta\Pi_{A\tilde{A}}^{\mu\nu} =-i\ \delta_3\ (q^2 g^{\mu\nu}-q^\mu q^\nu) \ ,
\end{eqnarray}
which are precisely of the form required to cancel the infinities in Eq.~(\ref{eq:VPA}). In the minimal subtraction scheme we obtain
\begin{eqnarray}
\delta_3 = - \frac{e^2}{24\pi^2}\ \frac{1}{\epsilon}\ .
\label{eq:delta3}
\end{eqnarray}

\subsection{Self Energy Amplitudes}

\begin{figure}[!t]
\begin{center}
\includegraphics[width=5.2in,height=3.0in]{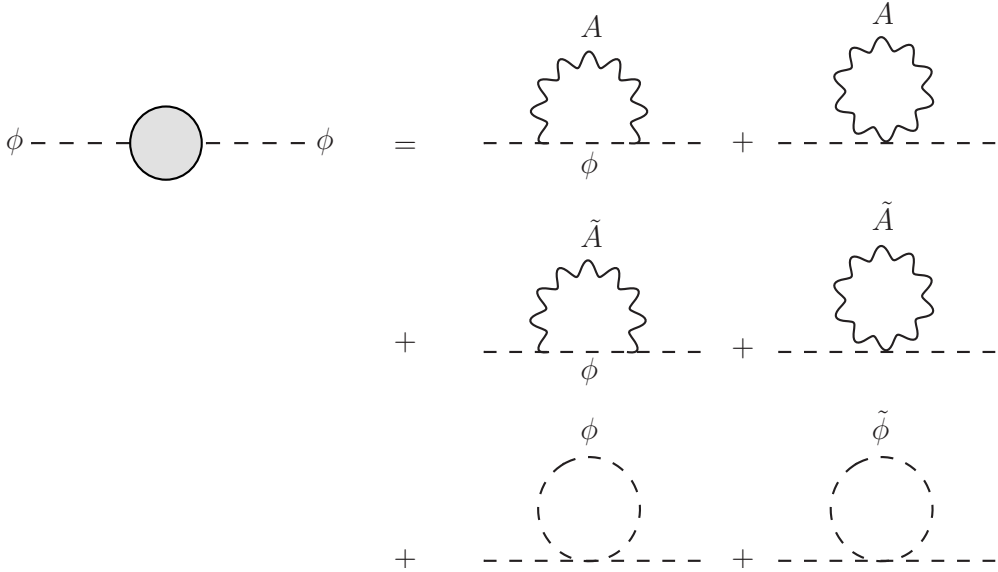}
\end{center}
\caption{One-loop contribution to the 1PI self energy amplitude $\Sigma_{\phi\phi}$ in the no-mixing $\xi=1$ gauge. In the ordinary gauge there are also diagrams involving internal mixed gauge propagators, $P_{A\tilde{A}}$.}
\label{fig:self}
\end{figure}

We will calculate the one loop contribution to the self energy amplitude $\Sigma$ for a scalar field in the no-mixing $\xi=1$ gauge and then will repeat the calculation in the ordinary $\xi=1$ gauge as a check. The relevant diagrams for the $\phi$ field, in the no-mixing gauge, are shown in Fig.~\ref{fig:self}. Those for the $\tilde{\phi}$ field are obtained by replacing $\phi$ with $\tilde{\phi}$; given the form of the Lagrangian (\ref{eq:Ltwofields2}), we expect that the contributions of the diagrams involving internal gauge bosons will change sign. The mixed self energy amplitude $\Sigma_{\phi\tilde{\phi}}$ explicitly breaks the $U(1)\times U(1)$ symmetry to diagonal $U(1)$, and must vanish in the limit $\lambda\to 0$; therefore, only the diagrams with scalar loops will contribute to $\Sigma_{\phi\tilde{\phi}}$.

We begin our calculation of $\Sigma_{\phi\phi}$ in the no-mixing $\xi=1$ gauge by considering potential quadratic divergences.  First, we examine the gauge-scalar diagrams on the top and middle lines of Fig.~\ref{fig:self}. The first two diagrams correspond to the gauge-sector contribution in ordinary scalar QED, which is quadratically divergent. That quadratic divergence is canceled by the $\tilde{A}$ diagrams, as we now demonstrate. The $SO(1,1)$ symmetry acting on the vector fields guarantees that: (i) the gauge-scalar-scalar and gauge-gauge-scalar-scalar couplings involving the photon and LW photon are identical (up to an unphysical minus sign in the gauge-scalar-scalar coupling\footnote{This minus sign comes from the $A_\mu-\tilde{A}_\mu$ dependence. However we can always redefine $\tilde{A}_\mu$ to $-\tilde{A}_\mu$, which turns $A_\mu-\tilde{A}_\mu$ into $A_\mu+\tilde{A}_\mu$. This, for example, is the convention adopted in Ref.~\cite{Grinstein:2007mp}.}), and (ii) the LW photon propagator has a minus sign, relative to the photon propagator. As a result, each diagram with an internal LW photon is opposite in sign to its counterpart with an ordinary photon, and in the UV (where the LW photon mass becomes irrelevant), there is an exact cancellation of the quadratic divergences.  Likewise, moving to the diagrams in the bottom row of Fig.~\ref{fig:self}, we recognize that the first diagram is familiar from the ordinary $\phi^4$ theory, and is of course quadratically divergent. The second diagram exactly cancels the quadratic divergence, as the $SO(1,1)$ symmetry acting on the scalar fields guarantees the equality of the $|\phi|^4$ and $|\phi|^2|\tilde{\phi}|^2$ couplings, as well as the negative sign in the $\tilde{\phi}$ propagator.

Having established that $\Sigma$ is free from quadratic divergences, we may proceed to complete the one-loop calculation in no-mixing $\xi=1$ gauge. In dimensional regularization, near $d=4$, the result is
\begin{eqnarray}
-i\ \Sigma_{\phi\phi} &=& -i\ \frac{\lambda(M^2-m^2)+3 g^2 M_A^2}{16\pi^2}\ \frac{1}{\epsilon}\ +\ {\rm finite\ terms}\ , \nonumber \\
-i\ \Sigma_{\tilde{\phi}\tilde{\phi}} &=&  
-i\ \frac{\lambda(M^2-m^2)-3 g^2 M_A^2}{16\pi^2}\ \frac{1}{\epsilon} \ +\ {\rm finite\ terms}\ , \nonumber \\
-i\ \Sigma_{\phi\tilde{\phi}} &=& i\ \frac{\lambda(M^2-m^2)}{16\pi^2}\ \frac{1}{\epsilon} \ +\ {\rm finite\ terms} \ .
\label{eq:self1}
\end{eqnarray}
Notice that there is mass renormalization but not wavefunction renormalization, as the $1/\epsilon$ coefficients are $q^2$ independent in this gauge.  Moreover, in the limit of exact $SO(1,1)\times SO(1,1)$ symmetry (where $M\to m$ and $M_A\to 0$) the self energy amplitudes vanish exactly in the no-mixing $\xi = 1$ gauge, because the LW and ordinary propagators are then of equal magnitude and opposite sign.  The theory is still not finite, because, for example, the vacuum polarization amplitudes do not vanish.

From  Eq.~(\ref{eq:Lct}) we find that the relevant counterterm contributions are of the form:
\begin{eqnarray}
-i\ \delta\Sigma_{\phi\phi} &=& i\ \delta_2\ q^2 -\ i\ \delta_{m^2}\ , \nonumber \\
-i\ \delta\Sigma_{\tilde{\phi}\tilde{\phi}} &=& -\ i\ \delta_2\ q^2 +\ i\ \delta_{M^2}\ , \nonumber \\
-i\ \delta\Sigma_{\phi\tilde{\phi}} &=& -\ i\ \delta_{mM}\ ,
\label{eq:self1ct}
\end{eqnarray}
and in the minimal subtraction scheme, we conclude:
\begin{eqnarray}
&&\delta_2 = 0 \ ,
\label{eq:delta2higher}
\end{eqnarray}
and
\begin{eqnarray}
&&\delta_{m^2} = -\frac{\lambda(M^2-m^2)+3 g^2 M_A^2}{16\pi^2}\ \frac{1}{\epsilon} \ , \nonumber \\
&&\delta_{M^2} = \frac{\lambda(M^2-m^2)-3 g^2 M_A^2}{16\pi^2}\ \frac{1}{\epsilon} \ , \nonumber \\
&&\delta_{mM} = \frac{\lambda(M^2-m^2)}{16\pi^2}\ \frac{1}{\epsilon} \ .
\label{eq:deltam}
\end{eqnarray}
Inserting these results into Eq.~(\ref{eq:ctrel}) yields the following expression for the mixing angle:
\begin{eqnarray}
\eta=-\frac{\lambda}{16\pi^2}\ \frac{1}{\epsilon}  \ .
\label{eq:varepsilon}
\end{eqnarray}
Note that this vanishes in the $\lambda\to 0$ limit, as expected:  for $\lambda\to 0$ the theory acquires a global $U(1)\times U(1)$ symmetry (with the diagonal $U(1)$ gauged) under which $\phi$ and $\tilde{\phi}$ rotate independently. This prevents $\phi$-$\tilde{\phi}$ mixing terms from being radiatively generated.

The calculation in the ordinary $\xi=1$ gauge proceeds somewhat differently because mixed gauge propagators ($P_{A\tilde{A}}$) are present.  Once their effects are included, the quadratic divergences still cancel among the diagrams involving internal gauge propagators.   A direct computation of the self-energy functions then yields:
\begin{eqnarray}
-i\ \Sigma_{\phi\phi} &=& 
-i\ \frac{\lambda(M^2-m^2)+g^2(3 M_A^2 + m^2 - q^2)}{16\pi^2}\ \frac{1}{\epsilon} + {\rm finite\ terms}\ , \nonumber \\
-i\ \Sigma_{\tilde{\phi}\tilde{\phi}} &=& 
-i\ \frac{\lambda(M^2-m^2)-g^2(3 M_A^2 + M^2 - q^2)}{16\pi^2}\ \frac{1}{\epsilon} + {\rm finite\ terms}\ , \nonumber \\
-i \Sigma_{\phi\tilde{\phi}} &=& i\ \frac{\lambda(M^2-m^2)}{16\pi^2}\ \frac{1}{\epsilon} + {\rm finite\ terms} \ .
\label{eq:self2}
\end{eqnarray}
In this gauge, both mass renormalization and wavefunction renormalization are present. The counterterm contributions are of course still given by Eq.~(\ref{eq:self1ct}), which have the right form to cancel the both the $q^2$-dependent and $q^2$-independent infinities in Eq.~(\ref{eq:self2}). In minimal subtraction scheme one obtains the relationships:
\begin{eqnarray}
&&\delta_{m^2} = -\frac{\lambda(M^2-m^2)+g^2(3 M_A^2+m^2)}{16\pi^2}\ \frac{1}{\epsilon} \ , \nonumber \\
&&\delta_{M^2} = \frac{\lambda(M^2-m^2)-g^2(3M_A^2+M^2)}{16\pi^2}\ \frac{1}{\epsilon} \ , \nonumber \\
&&\delta_{mM} = \frac{\lambda(M^2-m^2)}{16\pi^2}\ \frac{1}{\epsilon} \ ,
\end{eqnarray}
which lead to the same expression for the mixing angle as in the no-mixing $\xi=1$ gauge, Eq.~(\ref{eq:varepsilon}). However, this time, the wavefunction renormalization counterterm is
\begin{eqnarray}
\delta_2 = -\frac{g^2}{16\pi^2}\ \frac{1}{\epsilon} \ .
\label{eq:delta2lower}
\end{eqnarray}
which differs from the result in the other gauge.

Since universality of the $U(1)$ gauge coupling insures that the scalar field wavefunction renormalizations are always exactly compensated by the vertex corrections, as Eq.~(\ref{eq:ward}) shows explicitly, all that remains to show at one loop, is the cancellation of all $SO(1,1)$ breaking amplitudes in the $\phi^4$ sector.

\subsection{$\phi^4$ Vertex}

\begin{figure}[!t]
\begin{center}
\includegraphics[width=6.0in,height=5.0in]{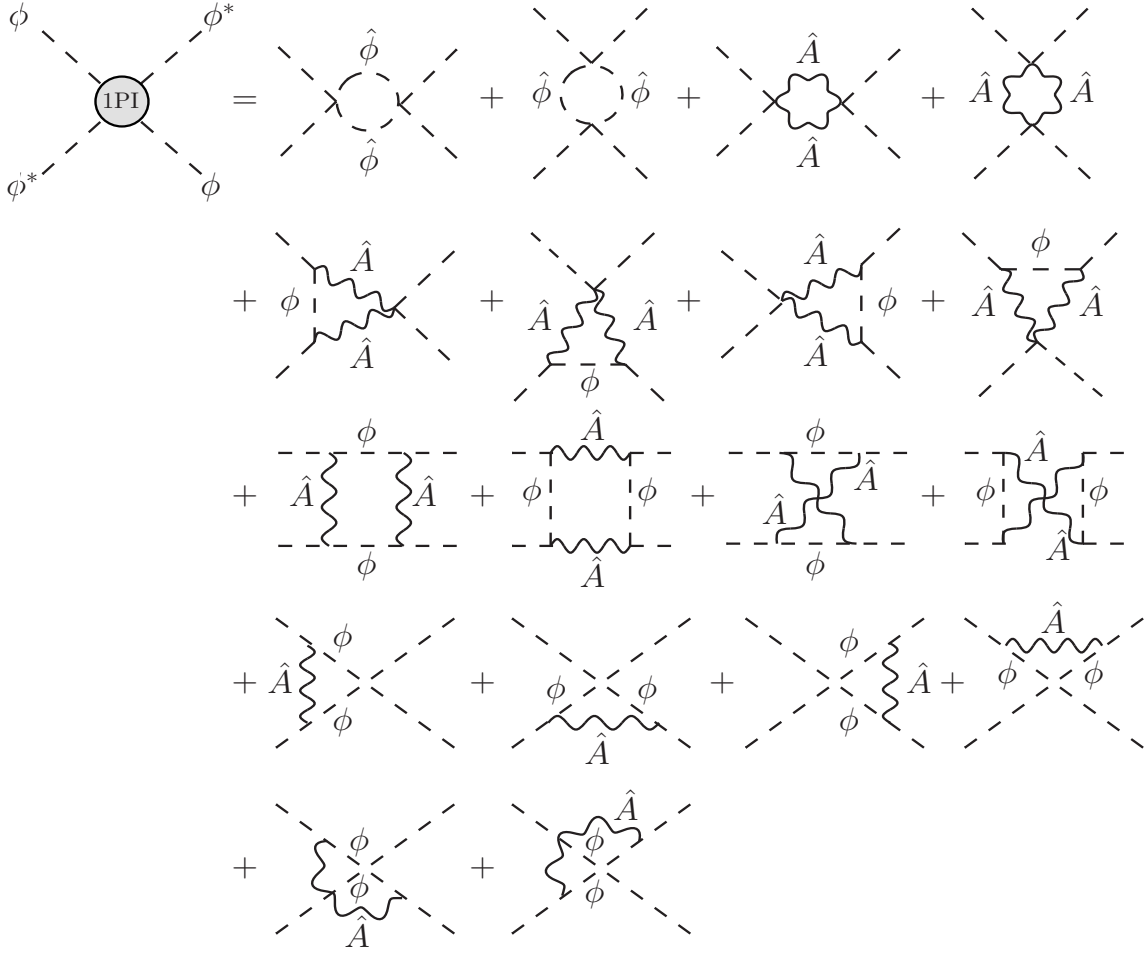}
\end{center}
\caption{One-loop contribution to the 1PI amplitude with four external $\phi$ fields. The number of diagrams is reduced by employing the hat field propagartors. See text for details.}
\label{fig:phi4}
\end{figure}

There are many diagrams contributing to the 1PI amplitudes with four external scalar fields. As with the self energy amplitudes, there are significant $SO(1,1)\times SO(1,1)$ cancellations involved. However, now that these are well understood, we can reduce the number of diagrams by using the hat-field propagators on all internal lines:  the $\hat{A}$ propagator of Eq.~(\ref{eq:hatprop}) when working in the ordinary $\xi=1$ gauge, the $\hat{A}$ propagator of Eq.~(\ref{eq:hatprophigher}) when employing the no-mixing $\xi=1$ gauge, and the gauge-independent $\hat{\phi}$ propagator  which is constructed by summing the simple propagators of the $\phi$ and $\tilde{\phi}$ fields:
\begin{eqnarray}
P_{\hat{\phi}\hat{\phi}} = P_{\phi\phi} + P_{\tilde{\phi}\tilde{\phi}} = \frac{i(M^2-m^2)}{(q^2-m^2)(M^2-q^2)} \ .
\end{eqnarray}
When we use the hat-field propagators, the expected cancellations occur within single diagrams, and are due to denominators with higher powers of loop momenta. 

Let us denote by $\Gamma_{f_1 f_2 f_3 f_4}$ the one-loop contribution to the 1PI amplitude with external scalar fields $f_1$, $f_2$, $f_3$, and $f_4$ and start by working within the no-mixing $\xi = 1$ gauge.  As a concrete example, we will consider $\Gamma_{\phi\phi\phi^\ast\phi^\ast}$, for which the relevant diagrams are those shown in Fig.~\ref{fig:phi4}. The first two diagrams are entirely due to the $\phi^4$ interaction, and involve only internal $\hat{\phi}$ fields. These diagrams are, thus, finite by power counting, since the $\hat{\phi}$ propagator decays like $q^{-4}$, at large momenta, and the vertices are momentum independent. The remaining two diagrams of the first row, as well as the diagrams of the second and third row are entirely due to the gauge-scalar interactions, which depend on $A_\mu-\tilde{A}_\mu\equiv \hat{A}_\mu$. Although the gauge-scalar-scalar vertices are momentum dependent, in the no-mixing gauge all these diagrams are finite by power counting. Finally, there are the diagrams in the last two rows, which involve both $\phi^4$ and gauge vertices. Once again these are finite by power counting in the no-mixing gauge.  The same reasoning holds for all of the $U(1) \times U(1)$ symmetric amplitudes; for the  $U(1)\times U(1)$ violating amplitudes $\Gamma_{\phi\phi\phi^\ast\tilde{\phi}^\ast}$, $\Gamma_{\phi\tilde{\phi}\tilde{\phi}^\ast\tilde{\phi}^\ast}$, and $\Gamma_{\phi\phi\tilde{\phi}^\ast\tilde{\phi}^\ast}$, the only difference is that the diagrams involving only gauge-scalar vertices do not contribute.  We conclude that the amplitudes are purely finite in the no-mixing gauge and 
\begin{eqnarray}
\delta_\lambda = 0 \ .
\label{eq:deltalambdahigher}
\end{eqnarray}

The situation is different in the ordinary $\xi=1$ gauge.  If we, again, start by considering $\Gamma_{\phi\phi\phi^\ast\phi^\ast}$, we still conclude that the first two diagrams, which involve only  the $\phi^4$ interaction, and involve only internal $\hat{\phi}$ fields are finite.  The ten diagrams involving only gauge-scalar interactions are another story.  Power-counting now  predicts a logarithmic divergence for each  of these diagrams, because of the $q^\mu q^\nu/M_A^2$ term in the $\hat{A}_\mu$ propagator.  However, we expect that these divergences must cancel against one another for symmetry reasons.  Recall that  the $U(1)\times U(1)$ violating amplitudes $\Gamma_{\phi\phi\phi^\ast\tilde{\phi}^\ast}$, $\Gamma_{\phi\tilde{\phi}\tilde{\phi}^\ast\tilde{\phi}^\ast}$, and $\Gamma_{\phi\phi\tilde{\phi}^\ast\tilde{\phi}^\ast}$ receive no contribution at all from the diagrams with only gauge vertices. Then those diagrams cannot make an infinite contribution to the $U(1)\times U(1)$-symmetric amplitudes like $\Gamma_{\phi\phi\phi^\ast\phi^\ast}$ either, since this would correspond to a hard breaking of the $SO(1,1)$ symmetry acting on the scalar fields. Explicit calculation confirms that the infinities arising from the diagrams with an even number of gauge vertices (last two diagrams of the first row, and diagrams of the third row) are precisely cancelled by the infinities from the diagrams with an odd number of gauge vertices (diagrams of the second row).  Therefore, the only possible infinite contribution to $\Gamma_{\phi\phi\phi^\ast\phi^\ast}$ and the other 1PI amplitudes in this gauge must arise from the diagrams in the last two rows, which involve both $\phi^4$ and gauge-scalar vertices.  Note that this implies that in the $\lambda\to 0$ limit all amplitudes with four external scalar fields are finite.
 
Computing the $\phi^4$ vertex correction diagrams from the last two rows of  Fig.~\ref{fig:phi4} in ordinary $\xi = 1$ gauge yields the following infinite contributions:
\begin{eqnarray}
&&i\ \Gamma_{\phi\phi\phi^\ast\phi^\ast} = i\ X + {\rm finite}\ , 
\quad i\ \Gamma_{\phi\tilde{\phi}\phi^\ast\tilde{\phi}^\ast} =  i\ 4 X + {\rm finite} \ , \quad
i\ \Gamma_{\tilde{\phi}\tilde{\phi}\tilde{\phi}^\ast\tilde{\phi}^\ast} = i\ X + {\rm finite} \ , \nonumber \\
&&i\ \Gamma_{\phi\phi\phi^\ast\tilde{\phi}^\ast} = -\ i\ 2 X + {\rm finite} \ , \quad 
i\ \Gamma_{\phi\tilde{\phi}\tilde{\phi}^\ast\tilde{\phi}^\ast} = -\ i\ 2 X + {\rm finite} \ , \quad
i\ \Gamma_{\phi\phi\tilde{\phi}^\ast\tilde{\phi}^\ast} = i\ X + {\rm finite} \ ,
\label{eq:1PI}
\end{eqnarray}
where
\begin{eqnarray}
X \equiv  -\frac{\lambda\ e^2}{8\pi^2}\ \frac{1}{\epsilon} \ .
\end{eqnarray}
At the same time, Eq.~(\ref{eq:Lct}) provides the counterterm contributions
\begin{eqnarray}
&&i\ \delta\Gamma_{\phi\phi\phi^\ast\phi^\ast} = -\ i\ \delta_\lambda \ , 
\quad i\ \delta\Gamma_{\phi\tilde{\phi}\phi^\ast\tilde{\phi}^\ast} = -\ i\ 4 \delta_\lambda  , \quad
i\ \delta\Gamma_{\tilde{\phi}\tilde{\phi}\tilde{\phi}^\ast\tilde{\phi}^\ast} = -\ i\ \delta_\lambda  \ , \nonumber \\
&&i\ \delta\Gamma_{\phi\phi\phi^\ast\tilde{\phi}^\ast} = \ i\ 2 \delta_\lambda  \ , \quad 
i\ \delta\Gamma_{\phi\tilde{\phi}\tilde{\phi}^\ast\tilde{\phi}^\ast} = \ i\ 2  \delta_\lambda  \ , \quad
i\ \delta\Gamma_{\phi\phi\tilde{\phi}^\ast\tilde{\phi}^\ast} =\ - i\ \delta_\lambda  \ .
\label{eq:1PIct}
\end{eqnarray}
which are precisely of the form required to cancel the infinities in Eq.~(\ref{eq:1PI}).  In the minimal subtraction scheme we obtain
\begin{eqnarray}
\delta_\lambda = - \frac{\lambda\ e^2}{8\pi^2}\ \frac{1}{\epsilon} \ .
\label{eq:deltalambdalower}
\end{eqnarray}

\subsection{Running of $g$ and $\lambda$}

We will now determine the $\beta$-functions of the LW theory and compare them with the results for ordinary scalar QED. To lowest order, the $\beta$-functions for the electromagnetic and $\phi^4$ couplings in the LW theory are given by
\begin{eqnarray}
\beta_g = \frac{g^2}{2}\,\mu\frac{\partial}{\partial\mu}\delta_3 \ , \qquad\qquad 
\beta_\lambda = \mu\frac{\partial}{\partial\mu}\left(-\delta_\lambda+2\ \lambda\ \delta_2\right) \ ,
\end{eqnarray}
where $\mu$ is the scale we must introduce in dimensional regularization to make the log arguments dimensionless, and $1/\epsilon$ in the counterterms is interpreted as:
\begin{eqnarray}
\frac{1}{\epsilon} \longrightarrow \log\frac{\Lambda^2}{\mu^2} \ .
\end{eqnarray}

Since we are employing a mass-independent renormalization scheme, below the LW mass scale ($M_{LW}$) we must impose the decoupling theorem and integrate out the LW fields.  Since what remains is identical to ordinary scalar QED, the counterterms are
\begin{eqnarray}
\delta_3 = - \frac{g^2}{48\pi^2}\ \frac{1}{\epsilon}\ , \quad\quad \delta_2 = \frac{g^2}{8\pi^2}\ \frac{1}{\epsilon} \ , \quad\quad
\delta_\lambda  = \left[\frac{\lambda^2}{8\pi^2} - \frac{\lambda\ g^2}{8\pi^2}\right] \frac{1}{\epsilon} \, .
\end{eqnarray}
These, in turn, yield the low-energy leading-order $\beta$-functions
\begin{eqnarray}
\beta_g (\mu < M_{\rm LW}) = \frac{g^3}{48\pi^2} \ , \quad\quad  \beta_\lambda (\mu < M_{\rm LW}) = \frac{\lambda^2}{4\pi^2} - \frac{3\ \lambda\ g^2}{4\pi^2}\ 
\end{eqnarray}
that are characteristic of ordinary scalar QED.

Above the LW mass scale, the appropriate counterterms are those we have derived for the full LW theory.   For the vector coupling, the counterterm value in Eq.~(\ref{eq:delta3}) leads to
\begin{eqnarray}
\beta_g (\mu > M_{\rm LW}) = \frac{g^3}{24\pi^2}~, 
\end{eqnarray}
which is twice the ordinary scalar QED $\beta_g$ function. In other words, the contribution from loops of the LW scalar is identical to that from loops of the ordinary scalar; since there are no internal gauge fields, the calculation is manifestly gauge invariant.  

For the $\beta_\lambda$ function above the LW scale we consider the no-mixing and ordinary gauges separately. If we employ the no-mixing $\xi=1$ gauge, then  Eqs.~(\ref{eq:delta2higher}) and (\ref{eq:deltalambdahigher}) tell us that $\delta_2$ and $\delta_\lambda$ are each separately zero. In this gauge the LW scalar and vector fields make contributions to the counterterms that are equal and opposite to those of the ordinary scalar and vector fields.  As a result, we find
\begin{eqnarray}
\beta_\lambda (\mu > M_{\rm LW}) = 0~. 
\end{eqnarray}
In the ordinary $\xi=1$ gauge, the values of $\delta_2$ are $\delta_\lambda$ are non-zero, as shown, respectively, by Eqs.~(\ref{eq:delta2lower}) and (\ref{eq:deltalambdalower}); however, the final result for $\beta_\lambda$ is the same, which provides a useful check of our calculations.

%%%%%%%%%%%%%%%%%%%%%%%%%%%%%%%%%%%%%%
%%%%%%%%%%%%%%%%%%%%%%%%%%%%%%%%%%%%%%
%%%%%%%%%%%%%%%%%%%%%%%%%%%%%%%%%%%%%%

\section{Discussion}\label{sec:conclusions}

In this paper we have discussed the global symmetries and the 
renormalizibility of Lee-Wick scalar 
QED. The combination of $SO(1,1)$ 
global symmetry, $U(1)$ gauge-invariance, and an  $SO(1,1)$ invariant gauge-fixing condition
  allow us 
to show directly  in the auxiliary-field formalism that the number of superficially divergent 
amplitudes in a LW Abelian gauge theory is finite.  To illustrate the renormalizability of
the theory, we have explicitly carried out the one-loop renormalization
program in LW scalar QED and demonstrated how the counterterms required are 
constrained by the joint conditions of gauge- and $SO(1,1)$-invariance. 
We have also computed the one-loop beta-functions
in LW scalar QED.

It would be interesting to generalize the discussion presented here to the case of non-Abelian
gauge theories. However, this is not immediately possible.
Notice that the $SO(1,1)$ transformation of Eq.~(\ref{eq:SOgauge}) mixes a gauge field, $A_\mu$, 
with a non gauge vector field, $\tilde{A}_\mu$. In an Abelian theory we have the freedom to 
promote $A^\prime_\mu \equiv\cosh\alpha\ A_\mu + \sinh\alpha\ \tilde{A}_\mu$ to a gauge field for 
two reasons. First, the requirement that $A^\prime_\mu-\tilde{A}^\prime_\mu$ transform like a 
gauge field gives us the freedom to choose which field should bear the transformation. Second, all 
gauge interactions depend solely on $e(A_\mu-\tilde{A}_\mu)$, which is an $SO(1,1)$ invariant. 
That these conditions are satisfied in Abelian gauge theory is perhaps not surprising, given that
a massive Abelian gauge theory is renormalizable \cite{Kroll:1967it}.

In a non-Abelian gauge theory, however, interactions do not depend solely on $g(A^a_\mu-\tilde{A}^a_\mu)$, and the $SO(1,1)$ symmetry is violated by the gauge interactions themselves. To see this consider the generalization of the gauge kinetic energy terms of Eq.~(\ref{eq:lower1}) to non-Abelian interactions,
\begin{eqnarray}
{\cal L}_{\rm gauge} = -\frac{1}{2}\ {\rm Tr}\ \hat{F}_{\mu\nu}^2 -2\ {\rm Tr}\  D^\mu\tilde{A}^\nu \hat{F}_{\mu\nu} ~,
%- M_A^2\ {\rm Tr}\ \tilde{A}_\mu^2 \ ,
\end{eqnarray}
where
\begin{eqnarray}
D^\mu\tilde{A}^\nu = \partial^\mu \tilde{A}^\nu - i g [\hat{A}^\mu,\tilde{A}^\nu] \ .
\end{eqnarray}
An $SO(1,1)$ transformation on the hat and tilde fields, and the gauge coupling, reads
\begin{eqnarray}
\hat{A}_\mu \to e^{-\alpha}\ \hat{A}_\mu \ , \quad \tilde{A}_\mu \to \sinh\alpha\ \hat{A}_\mu + e^\alpha\ \tilde{A}_\mu
\ ,\quad g \to e^\alpha\ g \ .
\end{eqnarray} 
Applying this to ${\cal L}_{\rm gauge}$ gives
\begin{eqnarray}
{\cal L}_{\rm gauge} \to {\cal L}_{\rm gauge} + i\ g\ \sinh\alpha\ e^{-\alpha}\ {\rm Tr}\ \hat{F}^{\mu\nu}[\hat{A}_\mu,\hat{A}_\nu] \ .
\label{eq:breaking}
\end{eqnarray}
Thus the $SO(1,1)$ symmetry associated with the vector fields is explicitly broken by dimension-four gauge-boson self-interactions.
 %Furthermore in non-Abelian gauge theories it is not clear whether a hard mass term for the vector bosons corresponds to a soft breaking, since typically this is associated to nonlinear sigma field kinetic terms\footnote{In Abelian gauge theories a hard mass term for the gauge field is truly soft, as this does not affect the renormalizability of the theory.}.

In principle we would therefore expect the $SO(1,1)$ breaking to propagate to other sectors of the theory, and spoil the special relations between couplings that guarantee the cancellation of quadratic divergences. However both 
power-counting in the
higher derivative formulation \cite{Grinstein:2007mp} and the high-energy behavior of
massive vector meson scattering in Lee-Wick gauge theory \cite{Grinstein:2007iz} suggest that the number
of superficially divergent diagrams remains finite and that non-Abelian LW gauge theory may be renormalizable.
%Perhaps the $SO(1,1)$ breaking operator on the right-hand side of Eq.~(\ref{eq:breaking}) does not induce additional symmetry breaking terms with infinite coefficient. Indeed this operator depends on the hat fields only, which have a better behavior in the ultraviolet. 
A more thorough understanding of non-Abelian LW gauge theories is therefore necessary in order to extend the results demonstrated here for Abelian theories.  
%We will address these issues in future work.

\acknowledgments

This work was supported in part by the US National Science Foundation under grants PHY-0354226 and PHY-
0854889. RSC and EHS gratefully acknowledge the Aspen Center for
Physics for its support while part of this work was completed. RF thanks D. Dietrich, M. Frandsen, C. Kouvaris, and M. Nardecchia for useful discussions.


\begin{thebibliography}{199}

%\cite{Grinstein:2007mp}
\bibitem{Grinstein:2007mp}
  B.~Grinstein, D.~O'Connell and M.~B.~Wise,
  %``The Lee-Wick standard model,''
  Phys.\ Rev.\  D {\bf 77}, 025012 (2008)
  [arXiv:0704.1845 [hep-ph]].
  %%CITATION = PHRVA,D77,025012;%%

%\cite{Lee:1969fy}
\bibitem{Lee:1969fy}
  T.~D.~Lee and G.~C.~Wick,
  %``Negative Metric and the Unitarity of the S Matrix,''
  Nucl.\ Phys.\  B {\bf 9}, 209 (1969).
  %%CITATION = NUPHA,B9,209;%%

%\cite{Lee:1970iw}
\bibitem{Lee:1970iw}
  T.~D.~Lee and G.~C.~Wick,
  %``Finite Theory of Quantum Electrodynamics,''
  Phys.\ Rev.\  D {\bf 2}, 1033 (1970).
  %%CITATION = PHRVA,D2,1033;%%

 %\cite{Carone:2009it}
\bibitem{Carone:2009it}
  C.~D.~Carone,
  %``Higher-Derivative Lee-Wick Unification,''
  Phys.\ Lett.\  B {\bf 677}, 306 (2009)
  [arXiv:0904.2359 [hep-ph]].
  %%CITATION = PHLTA,B677,306;%%

%\cite{Cutkosky:1969fq}
\bibitem{Cutkosky:1969fq}
  R.~E.~Cutkosky, P.~V.~Landshoff, D.~I.~Olive and J.~C.~Polkinghorne,
  %``A non-analytic S matrix,''
  Nucl.\ Phys.\  B {\bf 12}, 281 (1969).
  %%CITATION = NUPHA,B12,281;%%

%\cite{Grinstein:2008bg}
\bibitem{Grinstein:2008bg}
  B.~Grinstein, D.~O'Connell and M.~B.~Wise,
  %``Causality as an emergent macroscopic phenomenon: The Lee-Wick O(N) model,''
  Phys.\ Rev.\  D {\bf 79}, 105019 (2009)
  [arXiv:0805.2156 [hep-th]].
  %%CITATION = PHRVA,D79,105019;%%

%\cite{Jansen:1992xv}
%\bibitem{Jansen:1992xv}
%  K.~Jansen, J.~Kuti and C.~Liu,
%  %``The Triviality Higgs mass bound with higher derivative Lagrangian,''
%  Nucl.\ Phys.\ Proc.\ Suppl.\  {\bf 30}, 681 (1993).
%  %%CITATION = NUPHZ,30,681;%%

%\cite{Jansen:1992xx}
%\bibitem{Jansen:1992xx}
%  K.~Jansen, J.~Kuti and C.~Liu,
%  %``Strongly interacting Higgs sector in the minimal Standard Model?,''
%  arXiv:hep-lat/9212029.
%  %%CITATION = HEP-LAT/9212029;%%

%%\cite{Jansen:1993ji}
%\bibitem{Jansen:1993ji}
%  K.~Jansen, J.~Kuti and C.~Liu,
%  %``Strongly interacting Higgs sector in the minimal Standard Model?,''
%  Phys.\ Lett.\  B {\bf 309}, 127 (1993)
%  [arXiv:hep-lat/9305004].
%  %%CITATION = PHLTA,B309,127;%%

%\cite{Jansen:1993jj}
%\bibitem{Jansen:1993jj}
%  K.~Jansen, J.~Kuti and C.~Liu,
%  %``The Higgs model with a complex ghost pair,''
%  Phys.\ Lett.\  B {\bf 309}, 119 (1993)
%  [arXiv:hep-lat/9305003].
%  %%CITATION = PHLTA,B309,119;%%

%\cite{Rizzo:2007ae}
\bibitem{Rizzo:2007ae}
  T.~G.~Rizzo,
  %``Searching for Lee-Wick Gauge Bosons at the LHC,''
  JHEP {\bf 0706}, 070 (2007)
  [arXiv:0704.3458 [hep-ph]].
  %%CITATION = JHEPA,0706,070;%%
 
%\cite{Rizzo:2007nf}
\bibitem{Rizzo:2007nf}
  T.~G.~Rizzo,
  %``Unique Identification of Lee-Wick Gauge Bosons at Linear Colliders,''
  JHEP {\bf 0801}, 042 (2008)
  [arXiv:0712.1791 [hep-ph]].
  %%CITATION = JHEPA,0801,042;%%

%\cite{Alvarez:2009af}
\bibitem{Alvarez:2009af}
  E.~Alvarez, L.~Da Rold, C.~Schat and A.~Szynkman,
  %``Vertex Displacements for Acausal Particles: Testing the Lee-Wick Standard
  %Model at the LHC,''
  JHEP {\bf 0910}, 023 (2009)
  [arXiv:0908.2446 [hep-ph]].
  %%CITATION = JHEPA,0910,023;%%
 
%\cite{Krauss:2007bz}
\bibitem{Krauss:2007bz}
  F.~Krauss, T.~E.~J.~Underwood and R.~Zwicky,
  %``The Process $gg \to h(0) \to \gamma \gamma$ in the Lee-Wick standard
  %model,''
  Phys.\ Rev.\  D {\bf 77}, 015012 (2008)
  [arXiv:0709.4054 [hep-ph]].
  %%CITATION = PHRVA,D77,015012;%%
 

%\cite{Alvarez:2008za}
\bibitem{Alvarez:2008za}
  E.~Alvarez, L.~Da Rold, C.~Schat and A.~Szynkman,
  %``Electroweak precision constraints on the Lee-Wick Standard Model,''
  JHEP {\bf 0804}, 026 (2008)
  [arXiv:0802.1061 [hep-ph]].
  %%CITATION = JHEPA,0804,026;%%

%\cite{Underwood:2008cr}
\bibitem{Underwood:2008cr}
  T.~E.~J.~Underwood and R.~Zwicky,
  %``Electroweak Precision Data and the Lee-Wick Standard Model,''
  Phys.\ Rev.\  D {\bf 79}, 035016 (2009)
  [arXiv:0805.3296 [hep-ph]].
  %%CITATION = PHRVA,D79,035016;%%  

%\cite{Carone:2009nu}
\bibitem{Carone:2009nu}
  C.~D.~Carone and R.~Primulando,
  %``Constraints on the Lee-Wick Higgs Sector,''
  Phys.\ Rev.\  D {\bf 80}, 055020 (2009)
  [arXiv:0908.0342 [hep-ph]].
  %%CITATION = PHRVA,D80,055020;%%

%\cite{Dulaney:2007dx}
\bibitem{Dulaney:2007dx}
  T.~R.~Dulaney and M.~B.~Wise,
  %``Flavor Changing Neutral Currents in the Lee-Wick Standard Model,''
  Phys.\ Lett.\  B {\bf 658}, 230 (2008)
  [arXiv:0708.0567 [hep-ph]].
  %%CITATION = PHLTA,B658,230;%%
 
%\cite{Carone:2008bs}
\bibitem{Carone:2008bs}
  C.~D.~Carone and R.~F.~Lebed,
  %``Minimal Lee-Wick Extension of the Standard Model,''
  Phys.\ Lett.\  B {\bf 668}, 221 (2008)
  [arXiv:0806.4555 [hep-ph]].
  %%CITATION = PHLTA,B668,221;%%  
  
%\cite{Chivukula:2010nw}
\bibitem{Chivukula:2010nw}
  R.~S.~Chivukula, A.~Farzinnia, R.~Foadi and E.~H.~Simmons,
  %``Custodial Isospin Violation in the Lee-Wick Standard Model,''
  arXiv:1002.0343 [hep-ph].
  %%CITATION = ARXIV:1002.0343;%% 
  

%\cite{Grinstein:2007iz}
\bibitem{Grinstein:2007iz}
  B.~Grinstein, D.~O'Connell and M.~B.~Wise,
  %``Massive Vector Scattering in Lee-Wick Gauge Theory,''
  Phys.\ Rev.\  D {\bf 77}, 065010 (2008)
  [arXiv:0710.5528 [hep-ph]].
  %%CITATION = PHRVA,D77,065010;%%
  
%\cite{Carone:2008iw}
\bibitem{Carone:2008iw}
  C.~D.~Carone and R.~F.~Lebed,
  %``A Higher-Derivative Lee-Wick Standard Model,''
  JHEP {\bf 0901}, 043 (2009)
  [arXiv:0811.4150 [hep-ph]].
  %%CITATION = JHEPA,0901,043;%%
 
     %\cite{Kroll:1967it}
\bibitem{Kroll:1967it}
  N.~M.~Kroll, T.~D.~Lee and B.~Zumino,
  %``Neutral Vector Mesons and the Hadronic Electromagnetic Current,''
  Phys.\ Rev.\  {\bf 157}, 1376 (1967).
  %%CITATION = PHRVA,157,1376;%%

    
    %\cite{Grinstein:2008qq}
\bibitem{Grinstein:2008qq}
  B.~Grinstein and D.~O'Connell,
  %``One-Loop Renormalization of Lee-Wick Gauge Theory,''
  Phys.\ Rev.\  D {\bf 78}, 105005 (2008)
  [arXiv:0801.4034 [hep-ph]].
  %%CITATION = PHRVA,D78,105005;%%


\end{thebibliography}
\end{document}